\documentclass[11pt]{article}
\usepackage[a4paper,hmarginratio=1:1,vmarginratio=2:3,totalwidth=15.6cm,
totalheight=21.1cm]{geometry}
\usepackage{bm,epstopdf,epsfig,amsmath,amssymb,amsfonts,latexsym,comment}
\usepackage[rflt]{floatflt}
\usepackage[font=md,captionskip=8pt]{subfig}

\renewcommand{\baselinestretch}{1.38}

\newcounter{oldcounter}
\addtocounter{equation}{1}
\begin{document}
 \begin{flushright}
OUTP-0910P\\
\end{flushright}

\thispagestyle{empty}

\vspace{2.5cm}

\begin{center}
{\Large {\bf 
Sommerfeld factor for arbitrary partial wave processes
}}
\vspace{1.cm}

\def\thefootnote{\fnsymbol{footnote}}
\setcounter{footnote}{0}

  {\bf S.~Cassel\footnote{{\bf e-mail:} s.cassel1@physics.ox.ac.uk}
  }\\
 
\vspace{0.5cm}
 {Rudolf Peierls Centre for Theoretical Physics, University of Oxford,\\
 1 Keble Road, Oxford OX1 3NP, United Kingdom.}
 \end{center}

\medskip

\begin{abstract}
\noindent
The Sommerfeld factor for arbitrary partial wave processes is derived in the non-relativistic limit. The s-wave and p-wave numerical results are presented for the case of Yukawa interactions. An approximate analytic expression is also found for the Sommerfeld factor of Yukawa interactions with arbitrary partial waves, which is exact in the Coulomb limit. It is demonstrated that this result is accurate to within 10\% for some common scenarios. The non s-wave Sommerfeld effect is determined to be significant, and can allow higher partial waves to dominate cross sections.
\end{abstract}

\bigskip
\bigskip

\def\thefootnote{\arabic{footnote}}
\setcounter{footnote}{0}
\newpage
\setcounter{page}{1}


\section{Introduction}\label{intro}

In systems that contain interacting degrees of freedom, resonances commonly develop which cannot be described by a perturbative approach. This is generally known as the Sommerfeld effect \cite{Sommerfeld}, and has been widely studied in the fields of chemistry, nuclear and condensed matter physics, and also in the dynamics of mechanical and electrical systems. In particle physics, effective theories are constructed to deal with bound states, but the main tool for evaluating physical quantities such as cross sections and decay rates is perturbation theory. It is then common to neglect non-perturbative contributions. This is usually a reasonable approximation for relativistic particles, but for slow moving particles the influence of non-perturbative scattering interactions before and/or after the main interaction event becomes very significant.

Recently, the importance of the Sommerfeld effect in dark matter annihilation has been pointed out in \cite{Hisano:2003ec,Hisano:2004ds,Hisano:2006nn,Cirelli:2007xd,j1,MarchRussell:2008tu,ArkaniHamed:2008qn} for properly calculating thermal relic densities, and determining signals for indirect dark matter searches. Other examples where the Sommerfeld effect is crucial include threshold production of heavy states at colliders, and partial decay rates when the products have large phase space suppression. The purpose of this paper is to review the origin of the Sommerfeld effect from a particle physics perspective, and demonstrate its significance for some common interactions.

This work was initially motivated in order to determine the Sommerfeld factor for arbitrary partial wave ($l\neq 0$) processes, which had been overlooked. It is found that, even in cases where higher partial waves can safely be neglected at the perturbative level, these channels can become dominant when the non-perturbative physics is properly accounted for. For an application of the numerical p-wave results in this paper to relic density calculations, see~\cite{Cassel:2009pu}. During preparation of this paper\footnote{The numerical Yukawa p-wave results were first presented by the author in a talk at the 46th~International School of Subnuclear Physics, Erice, Sicily, 29~August - 7~September 2008 [http://www.ccsem.infn.it/issp2008].}, a similar derivation of the general Sommerfeld factor has been provided in~\cite{Iengo:2009ni,Iengo:2009xf}. 

Apart from for pure electromagnetic interactions, it is necessary to use numerical simulations in order to evaluate the Sommerfeld factor. The computational requirements can then be an obstacle to including the Sommerfeld effect in relevant calculations. In this paper, an approximate analytic expression is also found for the Sommerfeld factor of Yukawa interactions for arbitrary partial waves, which is exact in the Coulomb limit. So far as the author is aware this is a new result, and is found to be accurate to within 10\% for the most common applications.

The structure of the paper is as follows. Section~\ref{deriv} outlines the derivation of the Sommerfeld factor for arbitrary partial wave processes. Section~\ref{app} then determines the approximate analytic expression for the Sommerfeld factor in the presence of Yukawa interactions. In Section~\ref{results}, the numerical and analytic results are evaluated and compared. Section~\ref{conc} presents the conclusions.

\section{Derivation of the Sommerfeld factor}\label{deriv}

The non-perturbative physics that leads to the Sommerfeld effect can be thought of as the limit of perturbative Feynman diagrams with an infinite number of particle exchanges. For a two body incoming or outgoing state, it is useful to consider the non-perturbative 4-point vertex function, $\Gamma$, in order to quantify the Sommerfeld effect. In this section, the method for determining $\Gamma$ is first presented. This function is then used to relate the non-perturbative cross section of an arbitrary process to the perturbatively calculated prediction. The ratio of these results defines the ``Sommerfeld factor."

The non-perturbative 4-point vertex function is a solution of the Bethe-Salpeter equation:
\begin{figure*}[h!]
\center
\begin{minipage}{15.6cm}
\includegraphics[width=15.cm]{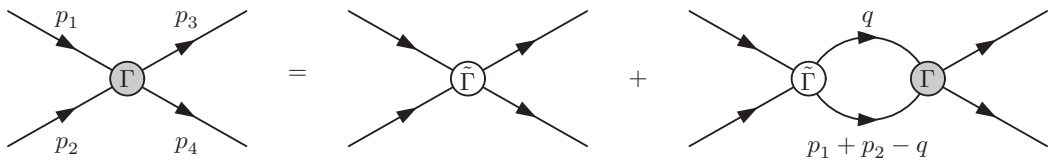} 
\end{minipage}
\caption{Bethe-Salpeter equation in diagrammatic form.}
\label{bs}
\end{figure*}

\vspace{-10mm}
\begin{eqnarray}
i \Gamma (p_1^{}, p_2^{}; p_3^{}, p_4^{} ) &=&
i \tilde{\Gamma} (p_1^{}, p_2^{}; p_3^{}, p_4^{} ) +
\int \frac{d^4 q}{(2 \pi)^4} ~ \tilde{\Gamma} ~ G(q) ~ G(p_1^{} + p_2^{} -q ) ~ \Gamma  
 \label{bse}
\end{eqnarray} 
where $\tilde{\Gamma}$ are ``compact" vertices which do not involve any intermediate state composed solely of the scattering particles (ie are described by two-particle irreducible diagrams), and $G$ is the non-perturbative propagator. The inhomogeneous integral equation for $\Gamma$ can be solved to arbitrary accuracy by a Liouville-Neumann series or Fourier methods, given that $\tilde{\Gamma}$ and the propagator are sufficiently well known from perturbative calculations. However, in the non-relativistic limit, such methods are difficult or sometimes impossible to implement. This case is most relevant for some common particle physics situations where the Sommerfeld effect is of great significance, and so is now concentrated on.

Fig~\ref{bs} can equivalently be expressed as the sum of Feynman diagrams with an increasing number of $\tilde{\Gamma}$ insertions. In the non-relativistic limit, $\tilde{\Gamma}$ usually is dominated by single particle exchange. For this reason, this class of diagram is often referred to as a ``ladder", (which can be better visualised when the scattering particle propagators in Fig~\ref{bs} are drawn as straight lines).

The approximation appropriate for the non-relativistic limit is that of instantaneous interactions \cite{Salpeter:1951sz}. This requires that the dependence of the momenta time components in the matrix element is removed. For the intermediate scattering particles that connect subsequent $\tilde{\Gamma}$ insertions, this prescription puts those propagators on-shell. The diagrams with finite ladders are then negligible compared to the infinite ladder diagram. The non-perturbative character of the vertex function $\Gamma$ is now manifest. In the non-relativistic limit, the Bethe-Salpeter equation becomes a homogeneous integral equation. To transform this equation into a more familiar form, it is convenient to introduce the function $\chi$ with the following definition:
\begin{eqnarray}
\chi (p_1^{} , p_2^{} ;p_3^{},p_4^{}) &=& G(p_1^{}) ~ G(p_2^{} ) ~ \Gamma (p_1^{} , p_2^{} ; p_3^{},p_4^{} )
\end{eqnarray}
From now on, the ($;p_3^{},p_4^{}$) arguments will implicitly be assumed in $\chi$. They do not influence the following steps for solving the Bethe-Salpeter equation, which is now expressed as:
\begin{eqnarray}
i   \chi (p_1^{}, p_2^{}) &\approx&  G (p_1^{}) ~G (p_2^{})
\int \frac{d^4 q}{(2 \pi)^4} ~ \tilde{\Gamma} (p_1^{},p_2^{} ; q,p_1^{}+p_2^{}-q) ~ \chi (q,p_1^{}+p_2^{}-q)
\end{eqnarray}
The leading contribution to $\tilde{\Gamma}$ in the non-relativistic limit is usually from single particle exchange. As an example, consider a scalar Yukawa theory with the interaction $g\,m_\phi^{} \, \phi^* \phi \, \varphi$, where $g$ is a dimensionless coupling constant. In the case of two $\phi$ particles scattering, the leading contribution to $\tilde{\Gamma}$ is:
\begin{eqnarray}
  \tilde{\Gamma} &\approx&  \frac{  -g^2 \, m_\phi^2  }{(p_1^{} - q)^2 - m_\varphi^2 }
\\
 \nonumber \\
&=&
    \frac{   g^2 \, m_\phi^2  }{|{\bf p_1^{} - q}|^2 + m_\varphi^2 - \omega^2 } 
\end{eqnarray}
where $\omega = (p_1^{})_0^{}-q_0^{}$ and the scattering interaction involves single $\varphi$ exchange. As mentioned before, the instantaneous approximation is required for the non-relativistic limit which sets $\omega =0$ in $\tilde{\Gamma}$. The perturbative vertex function then only depends on the transferred 3-momentum, $\tilde{\Gamma} =  U({\bf p_1^{} - q})$. This is a general feature in the non-relativistic limit, but it should be noted that the $\chi$ function still retains $q_0^{}$ dependence. The following parameters and function are now introduced:
\begin{eqnarray}
p ~=~ (p_1^{}-p_2^{})/2
\hspace{12mm}
P ~=~ (p_1^{}+p_2^{})/2
\hspace{12mm}
\tilde{\chi} (k_1^{},k_2^{}) ~=~ \chi (k_1^{}+k_2^{},k_1^{}-k_2^{})
\end{eqnarray}
In the centre of mass frame, $p=(p_0^{} ,{\bf p})$, and $P=(m_\phi^{} +E/2 \,,{\bf 0})$ where $E$ is the total kinetic energy of the system. The Bethe-Salpeter equation with the instantaneous approximation applied is then given by:
\begin{eqnarray}
i   \tilde{\chi} (P,p) &\approx&  G (P+p) ~G (P-p)
\int \frac{d^4 q}{(2 \pi)^4} ~ U({\bf p-q} ) ~ \tilde{\chi} (P,q-P)
\\
\nonumber \\
&=& G (P+p) ~G (P-p)
\int \frac{d^4 q^\prime}{(2 \pi)^4} ~ U({\bf p-q}^\prime ) ~ \tilde{\chi} (P,q^\prime)
\label{bsinst}
\end{eqnarray}
It is now useful to also introduce the Bethe-Salpeter wavefunction:
\begin{eqnarray}
\tilde{\psi}_{\mbox{\tiny BS}}^{} (\mathbf{q}) &=& \int \frac{d q_0^{}}{2\pi } ~\tilde{\chi} (P,q)
\end{eqnarray}
Using this and integrating eq~(\ref{bsinst}) over the $p_0^{}$ variable, the following equation is obtained for the scalar Yukawa case after re-arranging a multiplicative factor:
\begin{eqnarray}
\left( \frac{ {\bf p}^2}{m_{\phi} } -E \right) \, \tilde{\psi}_{\mbox{\tiny BS}}^{} (\mathbf{p}) +
\int \frac{d^3 q^\prime}{(2 \pi)^3} ~ V(\mathbf{p}-\mathbf{q}^\prime) ~  \tilde{\psi}_{\mbox{\tiny BS}}^{} (\mathbf{q}^\prime)
&=& 0
\end{eqnarray}
where $E$ is the total kinetic energy of the system, $U=-4m_\phi^2 \,V$, and only the leading term in $E/m_\phi^{}$ is kept. The Bethe-Salpeter wavefunction therefore is a solution of the Schr\"odinger equation (here in integral form), with a potential that accounts for the interactions included in $\tilde{\Gamma}$. For external states that have non-zero spin, the same steps can be applied and the Schr\"odinger equation is similarly found in the non-relativistic limit. For further discussion of the Bethe-Salpeter equation, see \cite{Salpeter:1951sz,Berestetsky:1982aq}.

The result of including the infinite series of scattering interactions has effectively transformed the incoming states from plane waves to a composite state described by $\tilde{\psi}_{\mbox{\tiny BS}}^{}$. The corrections to $\tilde{\Gamma}$ generally include annihilation channels. This introduces an imaginary term into the Sch\"odinger equation which results in a finite lifetime for the composite state. In scattering processes, bound states cannot be formed without radiating off energy. The composite state found then generally can not easily be described by an effective theory, and its influence is usually neglected.

For an example of including the non-perturbative scattering prior to the main interaction event, consider two body annihilation as shown in Fig~\ref{annpic}:
\begin{figure*}[h!]
\center
\includegraphics{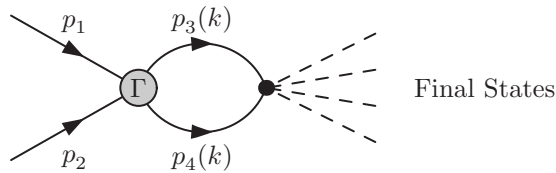} 
\caption{Diagram with non-perturbative scattering before annihilation}
\label{annpic}
\end{figure*}
\newline
The non-perturbative matrix element, ${\cal{M}}_{\mbox{\tiny with ladder}}^{}$, is related to the perturbative result, ${\cal{M}}_{\mbox{\tiny w/o}}^{}$, according to:
\begin{eqnarray}
{\cal{M}}_{\mbox{\tiny with ladder}}^{} (p_1^{} , p_2^{} ; \{p_f^{}\}) &=& 
\int \frac{d^4 k}{(2 \pi)^4} ~{\cal{M}}_{\mbox{\tiny w/o}}^{} (p_3^{}
, p_4^{} ; \{p_f^{}\}) \, G(p_3^{}) \, G(p_4^{}) \,
\Gamma(p_1^{},p_2^{} ; p_3^{}, p_4^{})
\hspace{3mm}
\label{mladd1}
\end{eqnarray}
where it is assumed that the Sommerfeld effect in the final states is negligible. This neglected feature can easily be included by the same technique, but for simplicity it is ignored here. The non-perturbative matrix element can be evaluated once the solution of $\Gamma$ in eq~(\ref{bse}) is found. In the non-relativistic limit, the instantaneous approximation should be applied to the perturbative matrix element, ${\cal{M}}_{\mbox{\tiny w/o}}^{}$. On performing the $k_0^{}$ integration for this case, it is seen that the Bethe-Salpeter wavefunction then weights the perturbative result as follows:
\begin{eqnarray}
{\cal{M}}_{\mbox{\tiny with ladder}}^{} (p_1^{} , p_2^{} ; \{p_f^{}\}) &\approx& 
\int \frac{d^3 k}{(2 \pi)^3} ~{\cal{M}}_{\mbox{\tiny w/o}}^{}
(\mathbf{k} ,
 - \mathbf{k} ; \{\mathbf{p}_f^{}\})
 ~\tilde{\psi}_{\mbox{\tiny BS}}^{} (\mathbf{k})
\label{mladd2}
\end{eqnarray}
where the centre of mass frame has been chosen. In a partial wave expansion, the leading terms in the non-relativistic limit for ${\cal{M}}_{\mbox{\tiny w/o}}^{}$, with a given angular momentum $l$, behave proportional to $k^l \, Y_{lm}^{}$ in the centre of mass frame, where $k$ here is the magnitude of the 3-momentum of either particle , and $Y_{lm}^{}$ is a spherical harmonic function.

For the s-wave case, it can immediately be seen by evaluating the integral that the non-perturbative matrix element differs by a factor of $\psi_{\mbox{\tiny BS}}^{} (\mathbf{r}=0)$ relative to the perturbatively calculated result, where $\psi_{\mbox{\tiny BS}}^{}$ is the position space representation of the Bethe-Salpeter wavefunction. The Sommerfeld factor for a given partial wave, $l$, is defined as:
\begin{eqnarray}
S_l^{} &=&
\frac{\mbox{non-perturbative partial wave cross section}}{\mbox{perturbative partial wave cross section}}
\label{sdef}
\end{eqnarray}
In the non-relativistic limit, $S_0^{}$ is simply $|\psi_{\mbox{\tiny BS}}^{} (\mathbf{r}=0)|^2$. In order to evaluate eq~(\ref{mladd2}) for higher partial waves, it is convenient to decompose the wavefunction in the (orthogonal) spherical harmonic basis. The wavefunction components can then also be written in a separable form as defined below:
\begin{eqnarray}
\tilde{\psi}_{Elm}^{} (\mathbf{k}) &=&  F_{El}^{} (k) \, Y_{lm} (\Omega_k^{} )
\label{sphh}
\end{eqnarray}
For a particular partial wave process, 
using ${\cal{M}}_{\mbox{\tiny w/o}}^{} = a_l^{} \, k^l \, Y_{lm}^{}$ in eq~(\ref{mladd2}) with $a_l^{}$ being some momentum independent factor, the non-perturbative matrix element is equivalently given by:
\begin{eqnarray}
{\cal{M}}_{\mbox{\tiny with ladder}}^{} &=&a_l^{}  \, \sum_{l^\prime, m^\prime}^{}
\int  \frac{d k}{(2\pi)^3} ~k^{l+2} F_{El^\prime}^{} (k)
\int d \Omega_k^{} \, Y_{lm}^{} (\Omega_k^{}) \, Y_{l^\prime m^\prime}^{*} (\Omega_k^{}) \, Y_{l^\prime m^\prime}^{} (\Omega_p^{})
\\
&=& a_l^{}  ~\sum_{l^\prime, m^\prime}^{}
\int \frac{d k }{2\pi^2}~k^{l+2} F_{El^\prime}^{} (k)
~ \delta_{l l^\prime}^{} ~\delta_{m m^\prime}^{} ~ Y_{l^\prime m^\prime}^{} (\Omega_p^{})
\label{msph}
\end{eqnarray}
where $\Omega_p^{}$ is the solid angle aligned with respect to the original collision axis of the incoming states. For a given partial wave, the Bethe-Salpeter wavefunction generated by scattering interactions is forced to have the same quantum numbers. In the perturbative case, the momentum appearing in the matrix element is fixed by the incoming states, $k = |{\bf p}_1^{}| = |{\bf p}_2^{}|$. The Sommerfeld factor can then be determined from eq~(\ref{msph}) using eq~(\ref{sdef}) to obtain:
\begin{eqnarray}
S_l^{} &=& \left|  \int \frac{d k}{2\pi^2} ~k^{l+2} \,F_{El}^{} (k)   ~ |{\bf p}_1^{}|^{-l} \right|^2
\label{slone}
\end{eqnarray}
In order to evaluate the result of integrating the momentum wavefunction weighted by different powers of $k$, it is useful to consider the radial Fourier transformation. This is now determined from the usual Fourier transformation:
\vspace{2mm}
\begin{eqnarray}
\psi_{Elm}^{} (\mathbf{r}) &=& \int \frac{d^3 k}{(2\pi)^3} ~\tilde{\psi}_{Elm}^{}
 (\mathbf{k}) ~ e^{-i \mathbf{k \cdotp r}} \nonumber \\
&=& \int \frac{k^2 dk}{(2\pi)^3}  \int d \Omega_k^{} ~ F_{El}^{} (k) \, Y_{lm}
 (\Omega_k^{} ) 
 ~  \sum_{l^\prime m^\prime} \,
  (-i)^{l^\prime}  \, j_{l^\prime}^{} (kr) \, Y_{l^\prime
 m^\prime}^{}
 (\Omega_r^{}) \, Y_{l^\prime m^\prime}^{*} (\Omega_k^{})
  \nonumber \\
&=&  \left[  (-i)^{l}  
 \int_0^\infty \frac{dk}{2 \pi^2} ~k^2 \, j_{l}^{} (kr) \, F_{El}^{} (k)  \right]  Y_{l m}^{} (\Omega_r^{})
 \nonumber \\[9pt]
&\equiv&   R_{El}^{} (r) ~Y_{l m}^{} (\Omega_r^{})
\label{four}
\end{eqnarray}
The result of differentiation of the radial wavefunction leads to the relation:
\begin{eqnarray}
\frac{\partial^{n} R_{E l}^{} (r)}{\partial r^n } &=& (-i)^l  \int_0^\infty
\frac{dk}{2\pi^2}~
 k^2
  \left[ \frac{ \partial^n j_l^{} (kr) }{\partial r^n} \right] F_{E l}^{} (k) 
\end{eqnarray}
A series expansion of the Bessel function then gives, $j_l^{}(kr) = (kr)^l / (2l+1)!! + O(kr)^{l+2}$ where $(2l+1)!! = (2l+1)! / (2^l \, l!)$, and so the following result is found:
\begin{eqnarray}
\frac{\partial^{l} R_{E l}^{}(r)}{\partial r^{l}} \Bigg|_{r=0}
 &=& 
 \frac{ l! }{(2l+1)!!} ~(-i)^l
 \int_0^\infty \frac{dk}{2\pi^2}~k^{l+2} \, F_{E l}^{} (k) 
\label{derr}
\end{eqnarray}
Using this in eq~(\ref{slone}), the Sommerfeld factor for the partial wave cross section with angular momentum $l$ is equivalently given by:
\begin{eqnarray}
S_{l}^{} &=&  \left|    \frac{  (2l+1)!!}{
  |{\bf p}_1^{}|^{l} ~ l!}  ~ \frac{\partial^{l} R_{E
 l}^{}(r)}{\partial r^{l}}
 \Bigg|_{r=0} \right|^2
\label{sfac}
\end{eqnarray}
where $|{\bf p}_1^{}|$ is the magnitude of the 3-momentum of either incoming particle in the centre of mass frame at infinite separation. This result has been derived in the non-relativistic limit. So far as the author is aware, this is a new result that, during preparation of this paper, has also been independently presented by Iengo \cite{Iengo:2009ni} (and see footnote 1).

\section{Analytic solutions for wavefunctions}\label{app}

In the previous section, it was shown that the Sommerfeld factor is related to solutions of the Sch\"odinger equation in the non-relativistic limit. The relevant potential is constructed by considering the interactions present in two-particle irreducible diagrams. For some common examples such as gauge boson, meson or Higgs interactions, a Yukawa potential is found at leading order. For a non-zero mass of the mediator responsible for the Yukawa interaction, there is no analytic solution for the wavefunction. This can then be determined numerically, however for many applications, an approximate analytic solution may be adequate for the level of precision desired. In this section, an approximate analytic wavefunction is found for the Yukawa potential and the Sommerfeld factor is then determined. The results are exact in the Coulomb limit, relevant for electromagnetic interactions.

Given a spherically symmetric potential, $V(r)$, the Schr\"odinger equation is separable. For a two-body system containing particles with a common mass $M$, the radial part of the wavefunction obeys the following:
\begin{eqnarray}
\left( - \frac{\, \hbar^2}{M } \, \partial_r^{2} - M \beta^2 + V(r) + \frac{\hbar^2 \, l(l+1)}{M r^2} \right) r \, R_l^{}(r) &=& 0 
\end{eqnarray}
where $\beta$ is the speed of each particle when at infinite separation in the centre of mass frame. Natural units ($\hbar = 1$) are now used for the rest of this discussion. For Yukawa interactions, the potential is of the form:
\begin{eqnarray}
V_Y^{} &=& - \frac{A\, e^{-m_\star^{} r }}{r}
\end{eqnarray}
where $m_\star^{}$ is the mass of the mediator for the interaction, and $A$ is the interaction strength which is positive (negative) for an attractive (repulsive) interaction. In order to find an approximate analytic solution, the following approximation is applied to the potential:
\begin{eqnarray}
V_Y^{} ~\sim ~V_H^{} &=& \frac{A \, \delta \, e^{-\delta r}}{1-e^{-\delta r}}
\end{eqnarray}
This approximation maintains the same short and long distance behaviour of the Yukawa potential, and is commonly known as the Hulth\'en potential. The question is then raised as to what choice of $\delta$ best reproduces the Yukawa potential. In order to answer this question for small strengths of the potential, parametrised by $A$, it is useful to consider the Schr\"odinger equation in an equivalent form known as the Lippmann-Schwinger equation:
\begin{eqnarray}
R_l^{} (r) &\propto & j_l^{} (kr) + M k \int_0^\infty r^{\prime \, 2} \, j_l^{} (k r_<^{}) \, n_l^{}  (k r_>^{}) \, V(r^\prime) \, R_l^{} (r^\prime) \, dr^\prime
\\
\nonumber \\
&\xrightarrow{~k,r\to 0\,}& \frac{(kr)^l}{(2l+1)!!} \left[ 1 -  \frac{M}{2l+1} \int_0^\infty r^\prime \, V(r^\prime) \, dr^\prime + O(A^2) \right]
\end{eqnarray}
where $r_<^{} = \mbox{min}(r,r^\prime)$, $r_>^{} = \mbox{max}(r,r^\prime)$ and $k=M\beta$. The Sommerfeld factor is sensitive to the behaviour of the wavefunction at the position space origin. In order for the approximation to be precise up to $O(A^1)$ in the limit of zero kinetic energy, it is necessary for the first moment of the radial potential, $\int_0^\infty r^\prime \, V (r^\prime) \, dr^\prime$, to be unchanged by the substitution $V_Y^{} \to V_H^{}$. This fixes the relation:
\begin{eqnarray}
\delta &=& \frac{\pi^2\, m_\star^{} }{6}
\label{del}
\end{eqnarray}
For $k \neq 0$, there is no analytic relation between $\delta$ and $m_\star^{}$ which equates the first order correction to the wavefunction, but limited progress can be made by performing a series expansion in $k/m_\star^{}$. In the following analysis, the identification in eq~(\ref{del}) is assumed for all strengths of the potential and kinetic energies. However, it should be stressed that more accurate results can be obtained if the choice of $\delta$ is improved for these cases.

The s-wave wavefunctions can now be found analytically, but for $l\neq 0$, a further approximation \cite{Greene} must be applied to the centrifugal term in order to permit analytic solutions:
\begin{eqnarray}
\tilde{V}_l^{}  &=& \frac{l (l+1)}{M} \frac{\delta^2 \, e^{-\delta r}}{(1-e^{-\delta r})^2}
\hspace{5mm} ~\approx~~ \frac{l (l+1)}{M r^2}
\hspace{3mm} \mbox{for $\delta r \ll 1$}
\label{vlapp}
\end{eqnarray}
This is reasonable for short range potentials, but does not reproduce the correct long distance behaviour. These approximations convert the effective potential of the radial Schr\"odinger equation into a form that allows solutions in terms of hypergeometric functions, and therefore is a special case of the Natanzon hypergeometric potential \cite{Natanzon:1979sr}. For the chosen approximations, $V_H^{} + \tilde{V}_l^{}\,$ is in the form also known as a Manning-Rosen or Eckart potential.

The physics of the Schr\"odinger equation is only sensitive to certain combinations of the free parameters. The following dimensionless variables are then constructed:
\begin{eqnarray}
w ~=~  \frac{y}{x} \, \frac{ m_\star   }{\delta } 
\hspace{15mm} x ~=~ \frac{A}{\beta}
\hspace{15mm} y ~=~ \frac{A M}{m_\star}
\hspace{15mm} z~=~ 2r M \beta
\end{eqnarray}
Introducing a further variable, $t=1-e^{-z/2w}$, the radial Sch\"odinger equation with the approximated Yukawa potential and centrifugal term takes the form:
\begin{eqnarray}
 \left(   \partial_t^{2} -  \frac{\partial_t^{}}{(1-t)} + \frac{w^2}{(1-t)^2}  + \frac{  w\,x - l (l+1)  }{ t (1-t)} -    \frac{ l  ( l+1)}{t^2}     \right) r(t) \, R_l^{} \left[ r(t) \right]
&=& 0
\end{eqnarray}
The regular solution of this differential equation is given by:
\begin{eqnarray} 
R_l^{}  &=&  \frac{t^{l+1} }{z  } ~\,  \frac{  e^{-iz/2}   }{ \Gamma(\lambda)} \,~\left| \frac{\Gamma (a^-) \, \Gamma (a^+)}{\Gamma ( 2i w)} \right|  ~ \, ^{}_2 F_1^{} (a^-, a^+; \, \lambda ;\, t)
\label{rsol}
\\[9pt]
a^\pm &=& 1+l + i w \left( 1 \pm  \sqrt{ 1 - x/w } \right)
\hspace{20mm}
\lambda ~=~ 2l+2
\end{eqnarray} 
where $_2^{} F_1^{}$ is a hypergeometric function which has the following behaviour at small $t$:
\begin{eqnarray*}
^{}_2 F_1^{} (a^+, a^-; \, 2\lambda ;\, t) &=& 1 + \frac{a^+ a^-}{2 \lambda \,} t + O(t^2)
\end{eqnarray*}
The normalisation of the wavefunction has been chosen such that an incident wave $e^{i{\bf k} . {\bf r}}$ of unit strength is obtained at infinity. This is then consistent with cross section calculations. In the spherical harmonic basis, an incident wave has the following decomposition:
\begin{eqnarray}
e^{i{\bf k} . {\bf r}} &\xrightarrow{~r\to \infty \,}& \sum_{l=0}^{\infty}  ~
\frac{1}{2ikr}  \left[ e^{ikr} - (-1)^l e^{-ikr} \right]  \,
\sum_{m=-l}^{l} Y^*_{lm} (\Omega_r^{}) \, Y_{lm}^{} (\Omega_k^{})
\end{eqnarray}
The limiting behaviour of the $_2^{} F_1^{}$ hypergeometric function is given below:
\begin{eqnarray}
_2^{} F_1^{} (a^- , a^+ ; \, \lambda ; \, t) &\xrightarrow{~ r \to \infty \,}&
\frac{\Gamma(\lambda) \Gamma(\lambda-a^- -a^+)}{\Gamma (\lambda - a^-) \Gamma (\lambda - a^+)} + e^{iz} 
\left( \frac{\Gamma(\lambda) \Gamma(a^- +a^+ - \lambda)}{\Gamma (a^-) \Gamma (a^+)} \right)
\end{eqnarray}
and so the constructed solution in eq~(\ref{rsol}) has the correct normalisation up to a phase factor\footnote{Note the behaviour of the $\Gamma$ function under complex conjugation; $\Gamma (h^*) = [\Gamma (h)]^*$.}, which is irrelevant for determining the Sommerfeld factor.
In the Coulomb limit ($w\to \infty$), the wavefunction reduces to:
\begin{eqnarray} 
R_l^{} \left[ w \to \infty \right] &=&  \frac{z^{l} \, e^{- i z /2}  }{ \Gamma(\lambda)}   ~~  e^{ \pi x/4  } ~\, \Gamma \left(1+l+\frac{ix}{2} \right) \, ^{}_1 F_1^{} \left(1+l+\frac{ix}{2};  \, \lambda ; \, i z \right)
\end{eqnarray} 
This result follows from eq~(\ref{rsol}) by considering the limiting behaviour of the $\Gamma$ function:
\begin{eqnarray}
\left| \Gamma (a+i b ) \right| &\xrightarrow{~ b \to \infty \,}& b^{a}  \,  e^{ - \pi b/2 }  \sqrt{\frac{2 \pi}{b} }  \,  \left[ 1+O\left( b^{-1} \right) \right] 
\hspace{5mm} \mbox{for} ~a,b \in \mathbb{R}
\label{glim}
\end{eqnarray}
where $\mathbb{R}$ is the set of real numbers, and the confluent hypergeometric function, $_1^{} F_1^{}$, is obtained using the identity:
\begin{eqnarray}
_1 F_1 (a_1^{};b;v) &=& \lim_{a_2^{} \to \infty}  \Big[ \, ^{}_2 F_1^{} (a_1^{},a_2^{};\, b;\, v/a_2^{}) \, \Big]
\label{flim}
\end{eqnarray}
The normalisation of the Coulomb wavefunction is inherited from eq~(\ref{rsol}), but a check of the limiting behaviour of the confluent hypergeometric function confirms that an incident wave of unit strength is found as $r\to \infty$.

The Sommerfeld factor in this dimensionless formalism is given by:
\begin{eqnarray}
S_{l,0}^{}  &=&  \left|   \frac{ (2l+1)!}{ (l!)^2 }  \, 
 \frac{\partial^{l} R_{l}^{}(z)}{\partial z^{l}} \Bigg|_{z=0} \right|^2
\end{eqnarray}
However, a naive application of the $l\neq0$ approximate Yukawa wavefunction in this formula can give results that violate the unitarity limit. This is a consequence of the approximation applied to the effective centrifugal potential. A more careful analysis is provided in the next section that properly accounts for the centrifugal approximation.

\subsection{Sommerfeld factor with a modified centrifugal term}

After modification of the centrifugal term, the free particle eigenstates are no longer plane wave solutions. An equivalent statement is that the canonical momentum of the position co-ordinate is modified by the approximation. The transformation between the position representation and canonical momentum representation is determined from the usual completeness relation:
\begin{eqnarray}
| \mathbf{r} \rangle &=& \int \frac{d^3 k^\prime}{(2\pi)^3} ~| \mathbf{k}^\prime \rangle \langle \mathbf{k}^\prime | \mathbf{r} \rangle
\label{complete}
\end{eqnarray}
Since the spherical harmonics are unaffected by the approximation, the same steps as in eq~(\ref{four}) can be applied to obtain the modified radial transformation relation:
\begin{eqnarray}
R_l^{} (r)  &=&   \int_0^\infty  \frac{dk^\prime}{2\pi^2} ~ k^{\prime \, 2} \, F_l^{} (k^\prime) \,R_{l}^{\mbox{\tiny (A=0)}} (k^\prime r) 
\label{rdmodi}
\end{eqnarray}
where $R_l^{\mbox{\tiny (A=0)}}$ is the free particle solution, which can be determined from eq~(\ref{rsol}) by taking the limit $x\to 0$:
\begin{eqnarray}
R_l^{\mbox{\tiny (A=0)}} (z) &=&  \frac{t^{l+1}  }{z  } ~\,  \frac{ e^{-iz/2} ~  l!   }{ \Gamma(\lambda) } \,~\left| \frac{  \Gamma \left(1+l + 2 i  w \right)}{\Gamma ( 2 i  w)} \right|  ~  ^{}_2 F_1^{} \left(1+l , 1+l + 2 i  w; \, \lambda ;\, t \right)
\hspace{10mm}
\\
\nonumber \\
&=& \frac{z^l ~l!}{\Gamma(\lambda)} ~ \left| \frac{ \Gamma \left(1+l + 2 i w \right) }{\, \Gamma ( 2i w) \, \left( 2w \right)^{ l+1} } \right| ~+ O\! \left( z^{l+1} \right)
\label{rser}
\end{eqnarray}
It follows from the above that the density of states remains unchanged, as assumed in eq~(\ref{complete}). 
In the limit $w\to \infty$, the correct centrifugal term is recovered and so it is expected that the standard radial Fourier transform, as given in eq~(\ref{four}), is also recovered. This result can be verified by noting the following identity:
\begin{eqnarray*}
j_l^{} (z/2) &=&  z^l ~  \frac{ e^{-iz/2} ~l! }{ \Gamma (\lambda)}  ~ _1^{} F_1^{} (1+l; \lambda ;iz)
\end{eqnarray*}
and considering the limiting behaviour of the $\Gamma$ function and $_2^{} F_1^{}$ function, given in eq~(\ref{glim}) and eq~(\ref{flim}). The standard transformation is indeed found in the limit $w\to \infty$.

The same derivation of the Sommerfeld factor as presented in Section~\ref{deriv} can be carried out up until eq~(\ref{slone}), where the Sommerfeld factor is given as an integral of the momentum space wavefunction weighted by various powers of momentum. At this point, to evaluate the integral, it is necessary to consider derivatives of eq~(\ref{rdmodi}) instead of the standard radial transformation. Using the series expansion in eq~(\ref{rser}) for the free particle solution, the Sommerfeld factor for a two body system with the modified centrifugal potential, $\tilde{S}_l^{}$, is then:
\begin{eqnarray}
\tilde{S}_{l}^{}  &=&  \left|   \frac{ (2l+1)!}{ (l!)^2 }  \, 
  ~ \frac{ \,  \Gamma ( 2i w) \, \left( 2w \right)^{ l+1} }{  \Gamma\! \left (1+l + 2 i w \right) }
~ \frac{\partial^{l} R_{l}^{}(z)}{\partial z^{l}} \Bigg|_{z=0} \right|^2
\end{eqnarray}
Note for $l=0$ and/or $w\to \infty$, this reduces to the usual Sommerfeld factor as expected when the centrifugal term and/or its substituted approximation vanishes. This can be seen using the relation $\Gamma (1+h) = h\, \Gamma (h)$ for the $l=0$ case, and eq~(\ref{glim}) for the $w\to \infty$ limit.

\subsection{Analytic Sommerfeld factor}

The approximate analytic Sommerfeld factor is now presented for the case of Yukawa interactions for arbitrary partial waves. Substituting the approximate wavefunction, given in eq~(\ref{rsol}), into the Sommerfeld factor which accounts for the modified centrifugal term, the following result is found:
\begin{eqnarray*}
\tilde{S}_{l}^{}  &=& 
\left|      
   \frac{   \, \Gamma (a^-) \, \Gamma (a^+) }{  \Gamma\! \left (1+l + 2 i w \right) }
    \, \frac{ 1 }{ l! }
 \right|^2
\end{eqnarray*}
In the Coulomb limit, this becomes:
\begin{eqnarray}
\tilde{S}_{l}^{}  &\xrightarrow{~ w \to \infty\, }&
\left|    \Gamma \! \left( 1+ l +\frac{ix}{2} \right)\!
 \right|^2 ~\frac{ e^{\pi x/2} }{ (l!)^2 }
\end{eqnarray}
Using the relation $\Gamma (1+h) = h\, \Gamma(h)$ recursively, and that $\left|\Gamma \left(1 + ib \right) \right| = \sqrt{\pi b \, \mbox{csch} (\pi b)}$ for real~$b$, the Coulomb case can equivalently be written as:
\begin{eqnarray}
S_{l \, >0}^{} &=& S_{0}^{}
\times \prod_{b=1}^{l} \left( 1+  \frac{ x^2}{4  b^2}  \right)
\hspace{10mm} \mbox{where} \hspace{3mm} S_{0}^{} ~=~ 
\frac{\pi x}{1 - e^{ - \pi x } }
\label{scoul}
\end{eqnarray}
Recall that this result is exact for the Coulomb limit. Although the perturbative cross sections of higher partial waves are suppressed by factors of $\beta^{2l}$ in the non-relavistic limit, this result indicates that in the zero velocity limit ($x\to \infty$), the Sommerfeld effect introduces a multiplicative factor that leads to all partial wave cross sections having the same velocity dependence.

In the large screening limit for the Yukawa potential ($w \to 0$), the Sommerfeld effect saturates when the de Broglie wavelengths of the scattering particles are much greater than the range of the potential. However, in the attractive case, resonances can develop when the system is close to bound states leading to large enhancements, and these states can form at zero energy. For finite range potentials such as the Yukawa case, the number of bound states present is always finite. Below a critical value for the coupling, no bound states can form and so the magnitude of the Sommerfeld factor is bounded. In the limit $w\to 0$, the Sommerfeld factor is approximately:
\begin{eqnarray}
\tilde{S}_{l}^{}  &\xrightarrow{~ w \to 0\, }&
 \left|  
   \Gamma \! \left(1+l+\sqrt{ wx } \, \right) \, \Gamma \! \left(1+l-\sqrt{ wx } \, \right) 
 \right|^2 / \, (l!)^4
 \label{sscrtr}
\end{eqnarray}
where $wx = y \, (m_\star^{} / \delta)$. 
The Gamma function has poles at 0 and the negative integers. Therefore, the above result suggests that resonances are found when $wx = (1+l+n)^2$ with $n$ being a non-negative integer. The critical values of $y$ for the exact and approximated Yukawa potentials that allow one bound state of angular momentum $l=0,1$ are given in Table~1. The resonance behaviour at these points will be demonstrated in Section~\ref{results}.
\begin{table}[!h]
\vspace{2mm}
\center
\begin{tabular}[c]{|c||c|c|}
\hline
Potential & $l=0$ & $l=1$
\\ \hline  
approx Yukawa & $\delta/\, m_\star^{}$ & $4\, \delta/\, m_\star^{}$
\\ \hline
exact Yukawa & 1.680 & 9.082
\\ \hline
\end{tabular}
\label{boundstates}
\def\baselinestretch{1.}
\caption{Critical values of $y~(= AM/m_\star^{})$ for one bound state with angular momentum $l$. In this paper, $\delta/m_\star^{} = \pi^2/6 ~(\approx1.645)$ has been used. The exact results are taken from \cite{Brau:2007zzd}.}
\end{table}

In the large screening limit, for small $wx$, the approximate analytic Sommerfeld factor behaves in a series expansion as:
\begin{eqnarray}
\tilde{S}_{l}^{}  &=& 1+2 \, wx \, \psi_1^{} (1+l) + O ( w, wx )^2
\label{sscr}
\end{eqnarray}
where $\psi_1^{}$ is the trigamma function, defined by:
\begin{eqnarray}
\psi_1^{} (h) &=& \frac{d^2}{dh^2}  \ln \Gamma (h)
\end{eqnarray}
For physical angular momentum, $\psi_1^{} (l+1) \sim 1/(l+ \frac12 )$ is a good approximation, exact in the large $l$ limit, and $18\% ~(3\%)$ accurate for $l=0 \,(1)$. This result suggests that for a given $l$, the magnitude of the deviation of the Sommerfeld factor from unity is roughly equivalent for the attractive and repulsive cases in this limit. The same behaviour is also seen in eq~(\ref{scoul}) for the small $x$ Coulomb limit, independent of $l$.

\section{Comparison of numerical and analytic results}\label{results}

In this section, the Sommerfeld factor determined from numerical simulations and using the approximate analytic result is presented and compared. The numerical results were found with the unapproximated Yukawa potential and centrifugal term. Fig~3 shows the Sommerfeld factor for the $l=0$ partial wave cross section.

The Coulomb limit is reproduced in the limit $y\to \infty$. For the attractive case, resonances are found with the critical points in agreement Table~1. In the repulsive case, there appears to be a qualitative symmetry between the large screening and Coulomb limits. This is expected by the similar forms that the approximated Sommerfeld factor takes in these limits. The transition between the two regions is roughly where the contours would intersect if the limiting results were extended. By equating the first order terms in a series expansion of eq~(\ref{scoul}) and~(\ref{sscr}), this is found at $y \sim (2l+1) x$.
\begin{figure}[t!]
\center
\subfloat[Attractive Potential]{\includegraphics[width=7cm] {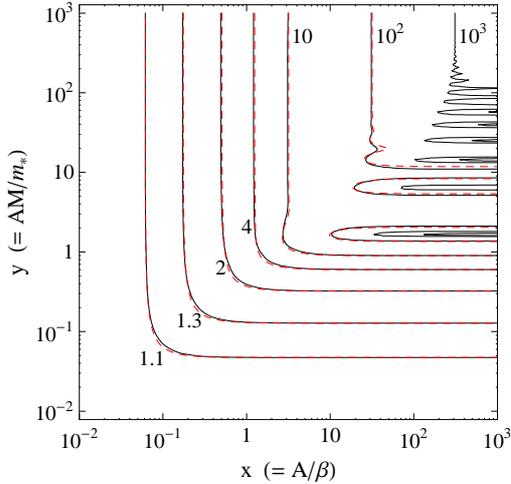}}
\hspace{10mm}
\subfloat[Repulsive Potential]{\includegraphics[width=7cm] {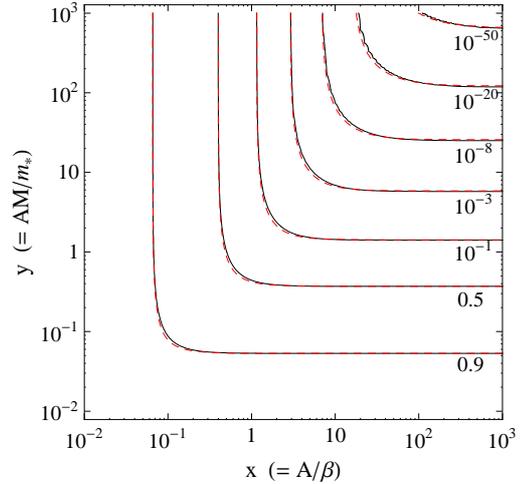}}
\label{somms}
\def\baselinestretch{1.}
\caption{
Sommerfeld Factor for a Yukawa interaction of a s-wave state
where the contours not labelled vary by a factor of 10. 
The solid lines represent $S_0^{}$ with the wavefunction determined from numerical simulations, and the red dashed lines are the approximate analytic result, $\tilde{S_0^{}}$. These contours are almost coincident except around the resonances where there is a noticeable shift. For clarity the contours of $\tilde{S_0^{}} \geq 10^3$ in (a)  are not shown.
}
\end{figure}

The analytic factor is accurate to within 10\% in the regions where $S_0^{} < 10$ and $S_0^{} > 0.1$, with the largest errors occurring when $y/x \sim 1$. This is not surprising as the choice of $\delta/m_\star^{}$ is only accurate in the limit $y/x \to 0$, and the exact Coulomb result is recovered as $y/x \to \infty$. The resonances are slightly off-position, with worse correlation for the higher resonances.

\begin{figure}[t!]
\center
\subfloat[Attractive Potential]{\includegraphics[width=7cm] {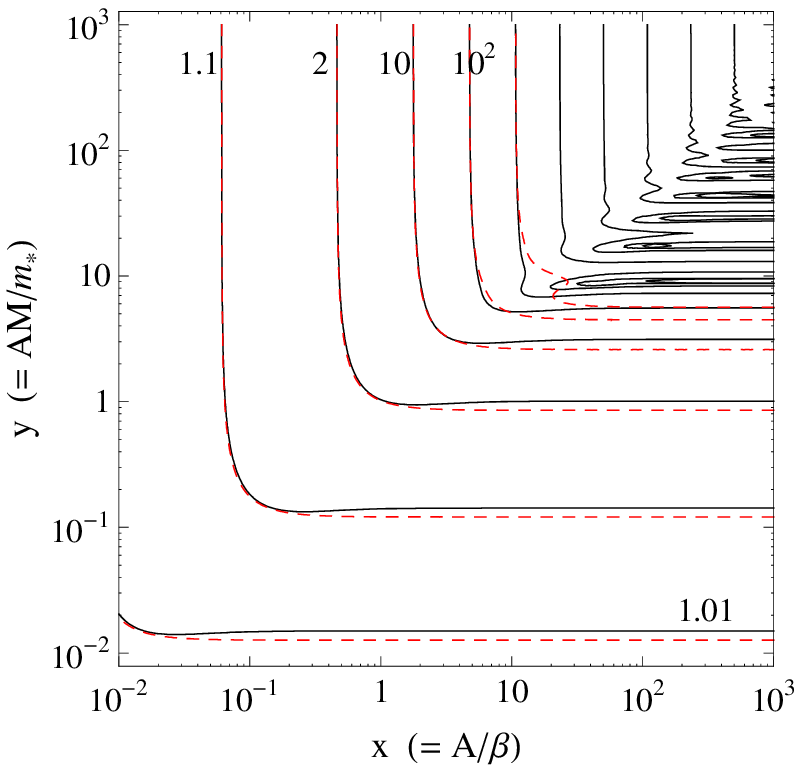}}
\hspace{10mm}
\subfloat[Repulsive Potential]{\includegraphics[width=7cm] {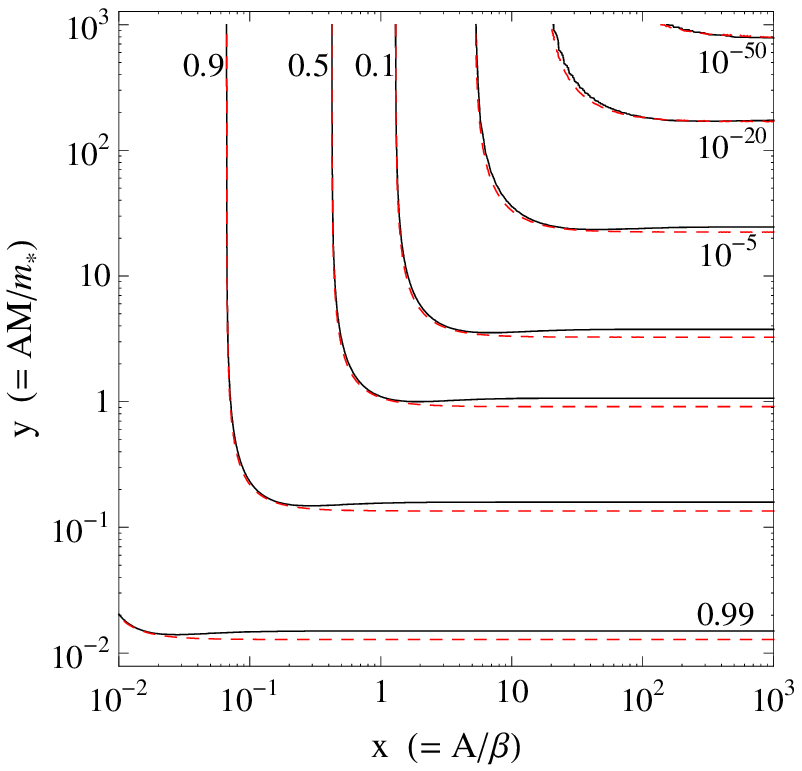}}
\label{sommp}
\def\baselinestretch{1.}
\caption{
Sommerfeld Factor for a Yukawa interaction of a p-wave state 
where the contours not labelled vary by a factor of 10. 
The solid lines represent $S_1^{}$ with the wavefunction determined from numerical simulations, and the red dashed lines show the approximate analytic result, $\tilde{S_1^{}}$. The contours are almost coincident in the region $y/x \gtrsim 1$ but have a mismatch in the region $y/x \lesssim 1$. For clarity the contours of $\tilde{S_1^{}} \geq 10^4$ in (a) are not shown.
}
\end{figure}
Fig~\ref{sommp} shows the Sommerfeld factor for the $l=1$ partial wave cross section. The correlation is relatively poor in the small $y/x$ region. For $y/x \ll 1$ and $y =1$ or $3$, the error of the analytic factor in reproducing the numerical result is roughly $10\%$ and $50\%$ respectively. For the attractive case, the incorrect resonance positions for the approximated result lead to order of magnitude errors for $y\gtrsim 6$. However, in the repulsive case, order of magnitude errors only are found for $y\gtrsim 30$, with the analytic factor being always less than the numerical result. The approximate analytic factor is then of limited use when the Sommerfed effect is large, without finding a better approximation.

If the shift $\delta \to 1.2\, \delta$  is made, the contours at small $y/x$ are nearly coincident with the numerical result. Actually, when the numerical results are determined for the Hulth\'en potential with an exact centrifugal term, the contours very closely match the Yukawa case without this shift. The expected origin of the error in the analytic factor therefore is dominantly from the approximation applied to the centrifugal term. A criterion has not been found to better select $\delta$ once this approximation is applied, but the following guess at a substitution was determined to significantly improve the results in the $l=1$ case:
\begin{eqnarray*}
w &\to& w \left[ 1- c_1^{}  \tanh \left( c_2^{} / w \right)  \right]
\end{eqnarray*}
For the choice $(c_1^{},c_2^{}) = (1/6,1/9)$, the results then agree to within $5\%$ in the regions where $S_1^{} < 10$ and $S_1^{} > 0.1$. The largest fractional error for the approximated analytic result in these regions still occur at $y/x \sim 1$. This transformation has effectively shifted the y variable in the small $y/x$ region and leaves it unchanged in the large $y/x$ region. The position of the resonances is still slightly offset though leading to large order of magnitude differences between the results where these features are present.

For the attractive Yukawa potential, the resonances of different partial waves do not coincide. It is then possible for a system to be dominated by a higher partial wave if close to a large $l$ resonance. For the repulsive case in the large screening region, the Sommerfeld suppression is reduced for higher partial waves since the centrifugal barrier becomes more effective at keeping the wavefunction away from the interaction core. It is found using eq~(\ref{sscrtr}) that the reduced Sommerfeld suppression for a successive partial wave compensates for the perturbative $\beta^{2l}$ suppresion, $ \beta^2 \, (S_l^{} / S_{l-1}^{}) \geq 1$, when:
\begin{eqnarray}
l^2 &\lesssim& \frac{y \, \beta}{1-\beta}
~~\lesssim~~ A  \left( \frac{2l+1}{1-\beta} \right)
\end{eqnarray}
where some $O(1)$ factors have been neglected, and the second inequality follows since the result  only applies in the large screening region, $y \lesssim (2l+1)\, x$. To determine for a particular case how many partial waves are significant and which channel dominates, it is necessary to also consider the relative magnitudes of the velocity independent factors in the partial wave expansion. For an attractive Yukawa interaction in the small $y$ region, the larger angular momentum processes are enhanced less by the same argument, so higher partial wave terms which are negligible in the perturbative expansion stay negligible.
 
 As is suggested by the numerical simulations, and also by the approximate analytic results, the Sommerfeld factor tends to infinity as $\beta \to 0$, when approaching resonances of the system. In physical situations, this would not occur as the zero energy bound states generally have finite lifetimes. A more careful analysis is then required in the close neighbourhood of these resonances.

\section{Conclusions}\label{conc}

The Sommerfeld effect can bring large order of magnitude corrections to cross sections. In this paper, the Sommerfeld factor relating the non-perturbative and perturbative matrix elements has been derived for arbitrary partial wave channels. It is demonstrated that in the non-relativistic limit, the incoming states are transformed by scattering interactions to effectively form a composite state whose wavefunction is a solution of the relevant Schr\"odinger equation. This wavefunction weights the perturbative matrix element in an integration over momentum space to give the non-perturbative matrix element.

The Sommerfeld factor for Yukawa interactions has been determined in an approximate analytic form for arbitrary partial waves, and also evaluated by numerical simulations for the $l=0,1$ cases. The s-wave result is found to be accurate to within 10\% when the non-perturbative cross section is up to an order of magnitude different to the perturbative result. For the p-wave factor, a correction introduced to compensate for the centrifugal approximation allowed the analytic result to be accurate to within 5\% for the same case. This is then a promising approach for application of the Sommerfeld effect when computational resources are prohibitively restrictive. However, the analytic result is poor at reproducing the correct resonance structure for attractive potentials. If the resonance region is probed, it remains necessary to use numerical simulations in order to obtain accurate results. The results are exact though in the Coulomb limit. It was found for certain areas in the parameter space that higher partial waves can dominate cross sections, when a perturbative analysis suggests that they are negligible. It can therefore be critically important to include the Sommerfeld effect for the $l\neq0$ case. 

One consequence of the Sommerfeld effect is the possibility that the annihilation mechanism which controls dark matter freeze-out could be different to the dominant mechanism of present day annihilation in our galactic neighbourhood. This is of relevance to identifying potential signals for indirect dark matter detection. For this case, ``boost" factors are commonly required in order to find a detectable signal. The Sommerfeld factor of an attractive interaction could provide such an enhancement. This can also be present in addition to other mechanisms such as internal Brehmstrahlung events or dark matter density perturbations.

During preparation of this paper, a similar derivation of the general Sommerfeld factor and evaluation for the Coulomb case has been presented in \cite{Iengo:2009ni}, in agreement with the results in this paper. In \cite{Iengo:2009xf}, the s- and p-wave Sommerfeld factors were also evaluated numerically given an attractive Yukawa interaction for various slices in the parameter space. Those results are also in agreement with that presented in this paper.

\section*{Acknowledgements}

The author is supported by the UK Science and Technology Facilities Council, 
and would like to thank Graham Ross, Dumitru Ghilencea, John March-Russell and Stephen West for many useful discussions.
\vspace{10mm}


\end{document}